\documentclass[a4paper,11pt]{article}
\usepackage{jheppub} 
\usepackage{lineno}

\makeatletter
\gdef\@fpheader{}
\makeatother

\arxivnumber{2609.20950} 

\usepackage[utf8]{inputenc}

\usepackage{orcidlink} 
\usepackage{enumitem}
\usepackage[toc,page]{appendix}
\usepackage{graphicx, xcolor, color}
\usepackage{caption}
\usepackage{subcaption}
\usepackage{amsfonts}
\usepackage{amsmath, empheq, amssymb, braket}
\usepackage{amsthm}
\usepackage{hyperref}
\hypersetup{
	colorlinks=True, 
	linktocpage=True,     
	linkcolor= red!40!brown,  
	citecolor = red!40!brown, 
	filecolor = red!40!brown
}

\usepackage[bb=stix]{mathalpha}

\newcommand{\Nf}{\mathrm{N}_{\text{F}}}
\newcommand{\eps}{\varepsilon}

\renewcommand{\d}{\mathfrak{d}}

\newcommand{\norm}[1]{\left\Vert #1 \right\Vert}

\newcommand{\abs}[1]{\left\vert #1\right\vert}
\renewcommand{\frak}[1]{\mathfrak{#1}}

\title{Zero Temperature, Degenerate Fermion Stars}

\author[4]{S. Boatto
	\thanks{\text{Email:} boatto@matematica.ufrj.br}
	\orcidlink{https://orcid.org/0000-0001-5708-0281}}
\author[1,2]{M. B. Paranjape \thanks{\text{Email:} paranj@lps.umontreal.ca}
	\orcidlink{https://orcid.org/0000-0002-2806-6568}}
\author[1]{V. Pessanha \thanks{\text{Email:} victor.mendes.de.oliveira@umontreal.ca}
	\orcidlink{https://orcid.org/0009-0009-4543-9408}}
\author[3]{C. A. D. Zarro \thanks{\text{Email:} carlos.zarro@if.ufrj.br}
    \orcidlink{https://orcid.org/0000-0002-4357-5168}}

\affiliation[1]{Département de Physique, Groupe de Physique des Particules, Université de Montréal}
\affiliation[2]{Centre de recherche mathématiques and Institut Courtois, Université de Montréal}
\affiliation[3]{Instituto de Física, Universidade Federal do Rio de Janeiro}
\affiliation[4]{Departamento de Matemática Aplicada, Instituto de Matemática, Universidade Federal do Rio de Janeiro}
    
\dedicated{Dedicated to Naresh Dadhich \\ a respected colleague and mentor, who left us too early.}

\abstract{
        We study a zero temperature, degenerate, fermion gas with fermions of rest mass $m$, that only interacts with itself, gravitationally.   We describe a self-consistent novel approach to the formation of a degenerate, quantum mechanical, star.  In this approach, the fermions occupy the energy levels that are determined by the self consistent gravitational interactions, according to the Pauli exclusion principle.  Then the size of the highest occupied level determines the radius of the star.   In the Newtonian gravitational limit, the fermions feel a potential corresponding to a simple harmonic well that smoothly attaches in the exterior to the usual $1/r$ Newtonian gravitational potential.  We find analytically, solutions for stars with any radius and mass, parametrized by $m$ and $N_s$, the total number of fermions. In the Newtonian limit, we easily find that we can construct a star that is more compact than the Buchdahl bound and indeed can even be smaller than the Schwarzschild radius. Clearly, such cases are outside of the purview of the Newtonian limit. Therefore we expand our analysis to fermions in the general relativistic context, interacting gravitationally with a metric.  We consider the metric corresponding to the interior Schwarzschild solution smoothly connecting to the usual exterior Schwarzschild metric. Treating matter semi-classically, we recover the Buchdahl bound in a completely independent analysis.  We briefly discuss classical stability of the Newtonian solution, and its relation to Newtonian polytropic fluids.}

\begin{document}
	\maketitle
    \flushbottom
    
	\section{Introduction}

    When considering self-gravitating systems in General Relativity (GR), one quickly encounters the famous relation between the Schwarzschild radius $R_s$ and total radius $R$ of the system, called the \textit{Buchdahl bound} \cite{buchdahl1959general}. It states that for spherically symmetric matter with outward non-increasing density and non-negative pressure anisotropy\footnote{Radial pressure is greater than or equal to tangential pressure}

    \begin{equation}
        \frac{R}{R_s} \geq \frac{9}{8}.
    \end{equation}

    The above is usually written in terms of the \textit{degree of compactification} $Z = \frac{R_s}{2R}$ which obeys the following upper bound.

    \begin{equation}
        Z \leq \frac{4}{9}.
    \end{equation}

    The bound can be derived from a purely geometric stand point \cite{Wald1984General, dadhich2020buchdahl, dadhich2025buchdahl} and it does not directly relate to matter stability, for such a study generally requires a separate analysis based on the specific matter content. It admits generalizations within the framework of classical matter and is actually modified if one considers cosmological constant, different topologies, extra dimensions, anisotropic/exotic matter with different equations of state or simply extend GR to other modified theories of gravity \cite{mazur2004gravitational, andreasson2008sharp, Zarro:2009gd, Zarro:2009zz, michael2025}.

    Gravitating systems composed of purely bosonic matter have also been studied within classical and semi-classical gravity \cite{jetzer1992boson, alcubierre2023boson, liebling2023dynamical, bovskovic2022soliton}, and also in the context of bosonic dark matter \cite{eby2016boson}, with the Buchdahl bound being generally preserved even for suitable generalizations thereof \cite{alcubierre2022extreme}.

    The case for fermionic matter is usually first studied in the context of neutron stars within the general relativistic framework on the basis of the (Tolman-Oppenheimer-Volkof) TOV equation \cite{Tolman1939,OppenheimerVolkoff1939} (see also \cite{Wald1984General,hartle2021gravity}), used to define the system's boundary and internal pressure and density functions. In these cases, a \textit{fermionic equation of state} is assumed and a relation between pressure and density is what is needed to solve the TOV equation numerically or analytically. This reasoning has been recently applied to Ho\v{r}ava-Lifschtiz gravity where a possible violation on the Buchdahl bound was found \cite{son2026vanishing}.

    In our work, we shift our focus from an explicit consideration of these intensive quantities and look to establish a relation between the mass and radius of the gravitating system in terms of the total number of particles and their corresponding masses in a self-consistent way, focusing strictly on the gravitational interaction between each constituent particle. The dynamical equation governing the quantum mechanical behavior of the particles is the Schrödinger equation with gravitational interactions. The zero temperature degenerate fermionic system is formed by filling the energy levels completely up to the Fermi energy.

    Our manuscript is divided into two parts: $i)$ Newtonian gravity with quantum fermionic matter and $ii)$ a generally covariant analysis with matter treated semi-classically. Section \ref{sec: setup} is dedicated to explaining the setup for these cases. In Section \ref{sec: weak gravity} we investigate the Newtonian gravitational case and in Section \ref{sec: semi classical gravity} the generally covariant one. Final remarks are given in Section \ref{sec: conclusions} and we explore a possible connection with Newtonian polytropic fluids in the Newtonian gravity case by means of a straight forward stability analysis argument left to Appendix \ref{sec: Matter stability}.

	\section{General Setup and Methods}\label{sec: setup}
	
	As mentioned above we are interested in self-consistently determining a relationship between the radius and the mass of our gravitating system of particles and seeing whether or not we find violations of the Buchdahl bound in either Newtonian or generally covariant treatments.
    
    In both cases, we consider a system of $N_s$ fermions with mass $m$ interacting purely gravitationally in spacetime. In the Newtonian case we define the system's radius $R$ by computing the average of the operator $\hat{R}^2 = \hat{x}^2+\hat{y}^2+\hat{z}^2$ over the last filled energy eigenstate found by solving Schrödinger's equation with a potential $V(r)$, defined on Section \ref{sec: weak gravity}, and then taking the square root. The total mass is found solving

    \begin{equation}
        M = N_s m + \frac{\mathrm{E}_\text{bind}}{c^2},
    \end{equation}
    in terms of $M$, with $\mathrm{E}_\text{bind}$ standing for the \textit{total gravitational binding energy} of the system, also defined in the next section. With $M$ and $R$ as functions of $N_s$ and $m$ we compute the degree of compactification $Z$ and check for Buchdahl bound violations.

    The generally covariant case on the other hand is realized in terms of the WKB approximation within curved spacetime. We make use of the \textit{interior Schwarschild solution}\footnote{The perfect fluid solution to Einstein's field equations smoothly connecting to the exterior, vacuum Schwarschild solution} (ISS) \cite{Wald1984General} and of the Klein-Gordon equation assuming small accelerations to derive a modified Schrödinger equation in the curved space time defined by the ISS. We are then able to rewrite the latter in terms of a momentum function $K(r)$, later used in the Bohr-Sommerfeld quantization procedure, where a semi-classical treatment of matter enables us to recover the Buchdahl bound to leading order in $\hbar$.
	
	\section{Newtonian gravity approximation}\label{sec: weak gravity}
	
	For this first scenario we assume the particle to be disposed in flat spacetime and to interact purely classically. We make this precise by introducing the gravitational potential $V(r)$ with which the particles will interact. We divide $V(r)$ into two spacial sub-regions, the one inside and the one outside the region defined by $R$.
	
	To make the definition for $V(r)$ inside sensible, we consider that the quantum star formed has a constant density $\rho = \frac{3M}{4\pi R^3}$. In this case, it is well known that the interaction potential will have to increase with the distance to the center square \cite{hartle2021gravity}. The definition for $V(r)$ outside on the other hand is simply given by the Newtonian potential generated by the mass $M$ of the quantum star. Thus
	
	\begin{equation}\label{eqn: QM potential}
		V(r) = 
		\begin{cases}
			-\frac{GMm}{r}, & r > R\\
			\frac{1}{2}\frac{mGMr^2}{R^3}-\frac{3GMm}{2R}, & r\leq R.
		\end{cases}
	\end{equation}
	
	We take the particles to be restricted to the interior region of the star and will thus be subject to interacting through its harmonic oscillator-like potential. 
	
	The interactions are non-relativistic and so, we can assume the wave functions of the particles to obey the time independent Schrödinger equation $\hat{H}\psi = E\psi$. Some algebraic manipulations get us to
	
	\begin{equation}\label{eqn: QM Hamilt eq - version one}
		-\frac{\hbar^2}{2m}\nabla^2\psi +\frac{1}{2}\frac{GMm}{R^3}r^2\psi = \tilde{E}\psi, \quad \tilde{E} = \frac{3GMm}{2R} + E .
	\end{equation}
	
	By comparing with the usual form of the harmonic oscillator potential, we are able to deduce that the frequency $\omega$ is
	
	\begin{equation}\label{eqn: omega freq equation}
		\omega = \sqrt{\frac{GM}{R^3}}.
	\end{equation}
	
	The energy levels are thus found to be
	
	\begin{equation}\label{eqn: tilde E expression}
		\tilde{E} = \hbar\omega\left(n_1+n_2+n_3 + \frac{3}{2}\right), \quad n_i\in \mathbb{Z}_{\geq 0} .
	\end{equation}
	
	An important observation is that, if we were interested in some particular quantity or observable that required specific knowledge of the wave functions of the system, we could not directly use the solutions of Eq.\eqref{eqn: QM Hamilt eq - version one}, for those represent \textit{bosonic} states, whereas we wish to consider a fermionic system. Instead, we would need to take the \textit{Slater determinant} and from there compute the quantity of interest.
	
	The fermionic behavior is henceforth \textit{imposed} in that we simply demand each particle occupy one energy level (indexed by $\frak{n} = (n_1, n_2, n_3)$) at a time. We will thus have one particle at $\frak{n} = (0,0,0)$, then another at either $\frak{n} = (1, 0, 0), (0, 1, 0)$ or $(0,0,1)$, and so on and so forth. This will be useful in the calculations to follow as we can make explicit use of annihilation and creation operators when computing the averages of certain operators.

	\subsection{Finding R and M}
	
	The radius of the system is defined by
    	
	\begin{equation}\label{eqn: R^2 from average of E_F}
		R^2 := \sum_{\norm{\frak{n}}_1 = N_F} \bra{\frak{n}} \hat{R}^2 \ket{\frak{n}},
	\end{equation}
    
    where $\hat{R}^2 = \hat{x}^2 + \hat{y}^2 + \hat{z}^2$ and $\Nf$ is the norm of $\frak{n}$ at the Fermi energy $\eps_F = \hbar \omega \left(N_F + \frac{3}{2}\right)$. Expanding $\hat{R}^2$ in terms of creation and annihilation operator will give us
	
	\begin{equation}
		\hat{R}^2 = \frac{\hbar}{m\omega}\left(\frac{3}{2} + \hat{n}_{x} + \hat{n}_{y} + \hat{n}_{z}\right) + \frac{\hbar}{2m\omega}\underset{k = x, y, z}{\sum} (\hat{a}_{k}^2 + (\hat{a}_{k}^\dagger)^2),
	\end{equation}
	
	from which Eq.\eqref{eqn: R^2 from average of E_F} is
	
	\begin{equation}\label{eqn: 1st equation for R}
		R^2 = \frac{\hbar}{m\omega} \left(N_F + \frac{3}{2}\right).
	\end{equation}
	
	Taking the square on both sides and using Eq.\eqref{eqn: omega freq equation} we end up with an expression relating $R$ and $M$, self-consistently determined from the coefficients appearing in the potential.
	
	\begin{equation}\label{eqn: R equation 1}
		R = \frac{\hbar^2}{GM m^2}\left(N_F + \frac{3}{2}\right)^2.
	\end{equation}
	
	For simplicity we will restrict our analysis to the leading order behavior of these quantities as function of $\Nf$ or, equivalently, the total number of particles $N_s$ which can be written as
	
	\begin{equation}\label{eqn: Ns explicit and assymptotic expressions}
		N_s = \underset{\norm{\frak{n}}_1\leq \Nf}{\sum} 1  = \frac{(\Nf+1)(\Nf+2)(\Nf+3)}{6} \simeq \frac{N_F^3}{6}.
	\end{equation}
	By putting together Eqs.(\ref{eqn: R equation 1}, \ref{eqn: Ns explicit and assymptotic expressions}) we get the following dependence on $N_s$ and $M$ for the total radius $R$
	
	\begin{equation}\label{eqn: R equation 2}
		R\simeq \frac{\hbar^2}{GM m^2}(6 N_s)^{\frac{2}{3}}. 
	\end{equation}
	
	To first order in $N_s$ we may use integration instead of summation to find the total binding energy $\mathrm{E}_\text{bind}$ as a function of $N_s, M$ and $R$. 
	
	\begin{equation}\label{eqn: binding energy - eq 1}
		\begin{split}
			\frac{\mathrm{E}_\text{bind}}{c^2} &= \frac{1}{c^2}\int_{\norm{\frak{n}}_1\leq \Nf} \left(\hbar\omega(n_x + n_y + n_z) - \frac{3GMm}{2R}\right) d^3n \\
			&= \frac{\hbar\omega N_s^{\frac{4}{3}}}{8c^2} - \frac{3GMm}{2Rc^2} N_s.
		\end{split}
	\end{equation}
	
	From Eq.\eqref{eqn: R equation 2} we may express the ratio $\frac{M}{R}$ and the oscillator frequency $\omega$ solely in terms of $M$. In doing so we conclude that
	
	\begin{equation}\label{eqn: binding energy - eq 2}
		\frac{\mathrm{E}_\text{bind}}{c^2} = -\frac{71}{48}\frac{m^3}{M_p^4}N_s^{\frac{1}{3}} M^2.
	\end{equation}
	
	By definition, the total mass $M$ of the system is given by the sum
	\begin{equation}
		M = N_s m + \frac{1}{c^2}\text{E}_\text{bind}.
	\end{equation}
	From Eq.\eqref{eqn: binding energy - eq 2}, elementary algebraic manipulations lead us to the following quadratic equation for $M$.
	
	\begin{equation}
		-\frac{71}{48}\frac{m^3}{M_p^4}N_s^{\frac{1}{3}} M^2 - M + N_sm =0,
	\end{equation}
	whose solution reads
	
	\begin{equation}\label{eqn: M total mass - final form}
		M = \frac{48 M_p^4}{71 m^3}N_s^{-\frac{1}{3}}\left( \sqrt{1 + \frac{71}{12}\frac{m^4}{M_p^4} N_s^{\frac{4}{3}}}-1\right).
	\end{equation}
    
	The corresponding expression for $R$ is found by plugging the above on Eq.\eqref{eqn: R equation 2}.
	
	\begin{equation}\label{eqn: R total radius - final form}
		R = \frac{71\sqrt[3]{36}}{48}\frac{Gm}{c^2}N_s\left(\sqrt{1 + \frac{71}{12}\frac{m^4}{M_p^4} N_s^{\frac{4}{3}}}-1\right)^{-1}.
	\end{equation}
	
	Note that once the mass $m$ of the particles is fixed, by taking $N_s\gg 1$, both $M$ and $R$ behave as $N_s^{\frac{1}{3}}$. This is interesting for the mass term $M$ in particular, as it signifies that the binding energy reduces greatly its growth with respect to $N_s$. Indeed, if we were to naively ignore it, we would have $M\sim N_s m$ in this limit, in which case Eq.\eqref{eqn: R equation 2} would have given us $R \sim \frac{\hbar^2}{Gm^2}N_s^{-\frac{1}{3}}$ instead. Moreover, if we consider the degree of compactification $Z$ defined by
    
    \begin{equation}\label{eqn: compactification parameter - definition}
        Z = \frac{GM}{c^2R} = \frac{R_s}{2R},
    \end{equation}
    with $R_s = \frac{2GM}{c^2}$ being the \textit{Schwarzschild radius}, we would observe that for a large enough number of particles, the radius of the star would decrease fast enough to make its degree of compactification unbounded, as here $Z  \sim \frac{m^4}{M_p^4}N_s^{\frac{4}{3}}$ in such a limit, meaning that the star could go arbitrarily deep inside its associated black hole.
    
    In a dynamical model, the usual interpretation is that this Newtonian approach (lacking the binding energy) simply fails at large $N_s$ and general relativistic corrections ought to be taken into account if we are to give a good description of how the compact object under consideration evolves.
	
	We should have in mind however that when $\frac{m}{M_p}$ is small when compared to $N_s^{-\frac{1}{3}}$, upon Taylor expanding the above solutions we exactly find that $M\sim N_s m$ and  $R \sim \frac{\hbar^2}{Gm^3}N_s^{-\frac{1}{3}}$. This suggests that the dimensionless quantity $\d = \frac{m}{M_p}N_s^{\frac{1}{3}}$ is the good physically meaningful parameter to have in mind.
	
	As a matter of fact, in the next subsection we will verify that from our current expressions for $M$ and $R$, we can violate the Buchdahl bound explicitly in what is the large $\d$ limit, and still keeping $Z$ finite, though greater than $\frac{1}{2}$.
	
	\subsection{Surpassing the Buchdahl bound}
	
	In the context of GR, for a matter distribution with positive outward non-increasing mass density $\rho(r)$, isotropic pressure $p(r)$, radius $R$ and ADM mass $M$ given by
	
	\begin{equation}
		M = \int_0^R4\pi r^2\rho(r)dr, 
	\end{equation} 
	a relationship between its Schwarzschild radius and $R$ has to be satisfied.
	
	\begin{equation}\label{eqn: BB}
		\frac{R}{R_s} \geq \frac{9}{8}.
	\end{equation}
	
	The above inequality is the aforementioned \textit{Buchdahl bound}. Although generally obeyed in its strict form, one can find objects for which the bound is saturated and equality holds \cite{boehmer2025buchdahl}, or even instances in which it is entirely modified by considering non-isotropic pressure fluids \cite{andreasson2008sharp}. The bound can also be reframed in terms of the degree of compactification
	
	\begin{equation}
		Z \leq \frac{4}{9},
	\end{equation}
	
	with equality representing the saturated case. It is clear that the Buchdahl bound provides a further constraint on the more trivial bound $Z\leq \frac{1}{2}$, stating that the star should lie outside the region bounded by its associated event horizon. 
	
	We now argue that our Newtonian model does present a violation of this inequality in a controlled way. That is, $Z$ achieves values grater than $\frac{4}{9}$, however those can not be arbitrarily large either.
	
	For this, we make use of Eqs.(\ref{eqn: M total mass - final form}, \ref{eqn: R total radius - final form}, \ref{eqn: compactification parameter - definition}) to find  
	
	\begin{equation}\label{eqn: compactification parameter - eq 1}
		Z  = \left(\frac{48 M_p^2}{71 m^2}\right)^2\frac{1}{\sqrt[3]{36} N_s^{\frac{4}{3}}}\left(\sqrt{1 + \frac{71}{12}\frac{m^4}{M_p^4} N_s^{\frac{4}{3}}}-1\right)^2.
	\end{equation}
	
	We can rewrite $Z$ as a function of the previously introduced dimensionless parameter $\frak{d} := \frac{m}{M_p}N_s^{\frac{1}{3}}$.
	
	\begin{equation}
		Z(\d) = \frac{1}{\sqrt[3]{36}}\left(\frac{48}{71}\right)^2\frac{1}{\d^4}\left(\sqrt{1 + \frac{71}{12}\d^4}-1\right)^2.
	\end{equation}
	
	It is easy to see that in the limit of small $\d$, $Z(\d)$ goes to zero. As $N_s$ is a natural number, this limit is achieved when $m$ approaches zero, in which case we interpret the star to be made of ultralight particles. On the other hand, we also find a finite positive value for the limit of big $\d$ which, for fixed $m$ means that the number of particles is huge, or for fixed $N_s$ that the star is made up of highly massive particles.
	
	\begin{equation}
		\lim_{\d\rightarrow\infty} Z(\d) = \frac{192}{71 \sqrt[3]{36}} \approx 0.8 .
	\end{equation}
	
	The fact that $Z$ is finite and greater than $\frac{1}{2}$ means that the final object lies inside the event horizon, with a radius strictly greater than zero. Although a mathematically quite interesting fact we ignore the viability of this process even within the framework of the current toy model, as highly compact objects even outside the event horizon cannot be described by Newtonian gravity and require strong gravitational corrections, besides other particles interactions happening on their inside, to be properly modeled.

	\section{General relativity and WKB approximation}\label{sec: semi classical gravity}
	
	We are now interested in extending the above construction to curved spacetime and see how much different are the semi-classical description of matter therein when compared to the previous analysis in the Newtonian gravitational case. For this, we make use of the Klein-Gordon (K-G) equation within the interior Schwarzschild solution (ISS) taken as our metric.
	
	\begin{equation}\label{eqn: ISS}
		ds^2 = - \left( \frac{3}{2}\sqrt{1 - \frac{R_s}{R}} - \frac{1}{2}\sqrt{1 - \frac{R_s r^2}{R^3}} \right)^2 c^2 dt^2 + \frac{dr^2}{1 - \frac{R_s r^2}{R^3} } + r^2 d\Omega^2.
	\end{equation}
    
	In analogy with the weak field limit approximation of GR we define a potential function per unit mass $\Phi(r)$ in terms of the $tt-$component of the above metric.
	
	\begin{equation}\label{eqn: Phi potential}
		g_{tt} =: -\left(1 + \frac{2\Phi}{c^2}\right).
	\end{equation}

	In the same spirit of the Newtonian gravitational case, it is worth recalling that the ISS is a solution to Einstein's field equations for an ideal fluid of constant density $\rho = \frac{3 M}{4 \pi R^3}$ and isotropic pressure that vanishes at the boundary, defined by $r = R$. We will not however be interested in such intensive quantities as the model itself is not dynamical, hence these should not play a role in the way we realize the system's description.

    The K-G equation in curved spacetime is given by

    \begin{equation}
        \frac{1}{\sqrt{\abs{g}}}\partial_\mu(\sqrt{\abs{g}}\partial^\mu \phi) - \frac{m^2 c^2}{\hbar^2}\phi = 0,
    \end{equation}
    where $\abs{g}$ is the absolute value of the determinant of the ISS metric (Eq.\eqref{eqn: ISS}). By making the substitution

    \begin{equation}
       \phi(t, x) = \psi(t, x) e^{-i\frac{mc^2}{\hbar}t}, 
    \end{equation}
    and by assuming a slowly varying condition on the function $\psi(t,x)$

    \begin{equation}
        \abs{\frac{\partial_t^2\psi}{c^2\left(1 + \frac{2\Phi}{c^2}\right)}}\ll 1,
    \end{equation}
    further algebraic manipulations yield us the following modified Schrödinger equation

    \begin{equation}\label{eqn: schr. equation in ISS}
        i\hbar\partial_t\psi = \frac{\hbar^2}{2m}\left(\psi'' + \psi'\left(\frac{2}{r} + \frac{1}{2}\sqrt{\abs{\frac{g_{rr}}{g_{tt}}}}\partial_r\left(\abs{\frac{g_{tt}}{g_{rr}}}\right)\right) + \frac{g_{rr}}{g_{tt}}\frac{2m^2}{\hbar^2}\left(g_{tt}\frac{\hbar^2}{2m^2r^2}\Delta_{\mathbb{S}^2} + \Phi\right)\psi   \right),
    \end{equation}
    with $\Delta_{\mathbb{S}^2}$ standing for the spherical Laplacian acting on $\psi$, and $\psi'$ is its derivative with respect to $r$. We can further simplify Eq.\eqref{eqn: schr. equation in ISS} by writing

    \begin{equation}
    \begin{split}
        P(r) &= \frac{2}{r} + \frac{1}{2}\abs{\frac{g_{rr}}{g_{tt}}}\partial_r\left(\abs{\frac{g_{tt}}{g_{rr}}}\right),\\[1.3mm]
        \psi(t, r, \theta, \varphi) &= u(r)e^{-\frac{1}{2}\int P(r)dr}\sum_{l, m}Y_{lm}(\theta, \varphi).
    \end{split}
    \end{equation}
    
    This transforms Eqn.\eqref{eqn: schr. equation in ISS} into the usual WKB form 

	\begin{equation}
		\frac{d^2 u}{dr^2} + K^2(r)u = 0,
	\end{equation}
    where $K(r)$ is the \textit{momentum function}, given by

	\begin{equation}\label{eqn: definition of K^2(r)}
		\begin{split}
			& K^2(r) = Q(r) - \frac{1}{2}\frac{dP}{dr} - \frac{P^2}{4},\\
			& Q(r) = \frac{2m^2}{\hbar^2}\frac{g_{rr}}{g_{tt}}\left(\Phi - g_{tt}\frac{\hbar^2}{2m^2r^2}l(l+1) - \frac{E}{m} \right).
		\end{split}
	\end{equation}

	The energy levels are quantized according to the Bohr-Sommerfeld quantization procedure, generally asserting that
	
	\begin{equation}\label{eqn: BS quantization condition - general form}
		\int_{r_1}^{r_2} K(r)dr = \pi\left(n + \frac{1}{2}\right),
	\end{equation}
	where $r_1$ and $r_2$ are the classical turning points found by solving $K(r) = 0$, and $n$ is the radial quantum number associated to the energy level considered.

    We now restrict our analysis to the Fermi level $\eps_F$, the highest occupied energy state. Before moving forward however, it should be clear that we are dealing with two viewpoints on the definition of the radius $R$ that we demand be equal. The first is to conceive it as the average of the operator $\hat{R}^2$ taken over the highest energy eigenstate for the system of $N_s$ particles, as analytically explored in the precedent section. The second is that it should be the point where the interior potential $\Phi(r)$ smoothly latches onto the \textit{exterior potential}\footnote{Defined analogously to $\Phi(r)$ on Eq.\eqref{eqn: Phi potential}, just with $\Phi_{\text{out}}(r)$ now coming from the exterior Schwarzschild solution instead.} $\Phi_{\text{out}}(r)$. This implies that $r = 0$ and $r = R$ have to be turning points for the function $K(r)$ in the semiclassical case. To first order in $\hbar$, the Fermi level is thus given by

	\begin{equation}\label{eqn: fermi energy formula - strong gravity}
		\eps_F = - mc^2\frac{R_s}{2R} + \mathcal{O}(\hbar^2),
	\end{equation}
	which makes the semiclassical expression for the momentum function become

    \begin{equation}
        K^2(r) = \frac{2m^2}{\hbar^2}\frac{g_{rr}}{g_{tt}}\left(\Phi + c^2\frac{R_s}{2R}\right) + \mathcal{O}(\hbar^0).
    \end{equation}
	
    The degeneracy of each energy state in taken into account by performing a similar calculation as the one in the preceding section, where $N_s \sim \frac{N_F^3}{6}$. 
    
    The fact that the origin becomes a turning point requires a change in Eq.\eqref{eqn: BS quantization condition - general form}, switching $N_F+\frac{1}{2}$ for $N_F+\frac{3}{4}$ \cite{brack2018semiclassical}. With all this taken into account, we can rewrite Eq.\eqref{eqn: BS quantization condition - general form} using the definition of $\Phi(r)$ coming from Eq.\eqref{eqn: Phi potential}.

	\begin{equation}\label{eqn: WKB approximation - large Ns integral}
		\begin{split}
			& \left(\frac{R^3}{R_s}\right)^{\frac{1}{2}}\sqrt{\beta} \int_1^{\frac{1}{\sqrt{\beta}}}\frac{1}{3-y}\sqrt{\frac{(y-1)(5-y)}{1-\beta y^2}} dy = \frac{\pi\hbar}{mc}\left((6 N_s)^{\frac{1}{3}} + \frac{3}{4}\right), \text{ where }\\[2mm]
			&\hspace{2.5cm} \beta = 1-\frac{R_s}{R}, \hspace{1mm} y = \frac{1}{\sqrt{\beta}}\sqrt{1-\frac{R_s}{R^3}r^2}.
		\end{split}
	\end{equation}

    In order to avoid integrating over a singularity, we must have

    \begin{equation}
        \frac{1}{\sqrt{\beta}} \leq 3 \Rightarrow \beta \geq \frac{1}{9}.
    \end{equation}

    The definition of the parameter $\beta$ tells us that the integral is finite and well defined if, and only if the Buchdahl bound is satisfied.

    \begin{equation}
        \frac{R_s}{R} \leq \frac{8}{9}.
    \end{equation}

    Furthermore, the treatment of the fermions does not allow us to explicitly know what the corresponding energy-momentum (EM) tensor would be. The ideal fluid EM tensor that the ISS gives us can only be thought of as an uncontrolled approximation to the actual EM tensor. 

    We conclude by conjecturing that a suitable accounting of the higher order terms in the $\hbar$ expansion would not yield significant changes to the bound on $\beta$, as the sub-leading order terms also depend on $\frac{1}{g_{tt}}$ which is the quantity responsible for the singularity appearing in the momentum function, and from which the Buchdahl bound is recovered.

	\section{Conclusions}\label{sec: conclusions}
	
	Our work concerned the study of completely degenerate quantum matter forming a quantum star within the frameworks of classical and semi-classical gravity. We took $N_s$ fermions of equal mass $m$ and assumed their interaction to be purely gravitational, modeling for instance the particle behavior of fermionic dark matter.
    
    We were interested in finding possible violations of the Buchdahl bound in these two regimes by self-consistently determining the total radius and total mass of the star formed by the system under the gravitational potentials considered.

    In particular, we approached the classical case by imposing the interaction to be that of a harmonic oscillator potential coming from the classical calculation of determining the gravitational potential of a uniform density, spherically symmetric body of mass $M$ and radius $R$. We solved Schrödinger's equation with such a potential and used the ladder operators to determine the average value of the operator $\hat{R}^2$, whose square root equals $R$, which is the size of the degenerate quantum system. The zero temperature fermionic behavior of the particles was imposed by completely filling all the energy levels up to the Fermi level. The total mass $M$ was then found by computing the total binding energy of this system $\mathrm{E}_\text{bind}$ (which is negative) and using the equation $M = N_s m + \frac{\mathrm{E}_\text{bind}}{c^2}$.

    Explicit expressions for $R$ and $M$ as functions of $N_s$ were found, which enabled an analytic description for the degree of compactification $Z$ as a function of the dimensionless variable $\d = \frac{m}{M_p}N_s^{\frac{1}{3}}$. We found for the classical case that the Buchdahl bound can be greatly surpassed and $Z$ is bounded above by around $0.8$, attained at large $\d$, meanwhile going to zero at small $\d$\footnote{Although at $Z = 0.5$ we are already at the Schwarzschild radius}.

    The semi-classical case was treated by using the Bohr-Sommerfeld quantization procedure within the WKB approximation scheme. We first took the interior Schwarschild solution (ISS), corresponding to the metric tensor of a constant density spherically symmetric mass with isotropic pressure, and there defined the gravitational potential $\Phi(r)$ in analogy to the weak field limit of GR in terms of the $tt-$component of such metric. 

    By imposing that the Klein-Gordon equation for $\phi(t, x) = \psi(t, x) e^{-i\frac{mc^2}{\hbar}}t$ on the ISS metric be such that the function $\psi$ has small accelerations, we were able to deduce a modified Schrödinger equation with potential $\Phi(r)$ in this curved geometry.

    We then moved to the time independent Schrödinger equation and used a standard Liouville transformation to put it in a form amenable to the WKB approximation, explicitly finding the momentum function $K(r)$ for this problem. We proceeded with a semi-classical limit of the equations by ignoring terms of order greater than one in $\hbar$ which lead to $r = 0$ and $r = R$ to be the classical turning points for $K(r)$. From the Bohr-Sommerfeld quantization formula we were able to recover the Buchdahl bound as a consistency condition, for the integral not to be done over a logarithmic divergence present in the momentum function.
	
	Lastly, we managed to use the usual stability analysis for polytropic fluids in the classical case to infer that our object is (classically) stable and could in principle be modeled as a Newtonian polytropic fluid with $\gamma = \frac{3}{2}$. As the quantum star is allowed to go beyond the horizon, as predicted by the upper bound on $Z$, a thorough study would be required to answer the question of whether or not it is possible to have a dynamically stable object after gravitational collapse. For this, besides a time-dependent metric and a relativistic framework, other particle interactions inside the star would have to be taken into account.

    \section*{Acknowledgments}
	
	We thank MITACS, the Agence Universitaire de la Francophonie, NSERC of Canada and the Département de Physique, Université de Montréal for financial support. S. B. and C. A. D. Z are partially supported by Conselho Nacional de Desenvolvimento Cient\'ifico e Tecnol\'ogico (CNPq) and Coordena\c{c}\~ao de Aperfei\c{c}oamento de Pessoal de N\'ivel Superior (CAPES) under finance code 001.  C. A. D. Z. is partially supported by CNPq under Grant No. 310703/2021-2 and by Funda\c{c}\~ao Carlos Chagas Filho de Amparo \`a Pesquisa do Estado do Rio de Janeiro (FAPERJ) under Grant No. E-26/201.447/2021. 
	
	We also thank Naresh Dadhich, discussions with whom  launched this investigation.

	
	\appendix
	\section{Matter stability in the Newtonian gravitational regime}\label{sec: Matter stability}
	
	For Newtonian polytropic fluids, whose equation of state can be summarized as
	
	\begin{equation}\label{eqn: NEwtonian EOS}
		p = K\rho^{\gamma},
	\end{equation}
	
	if we define $\gamma_\text{crit} = \frac{4}{3}$, it is well known that the stability condition for a compact object is that $\gamma >\gamma_\text{crit}$ \cite{weinberg1972gravitation}. The binding energy for a star whose matter content obeys Eq.\eqref{eqn: NEwtonian EOS} is found to be:
	
	\begin{equation}\label{eqn: Newtonian E_bind}
		\mathrm{E} = a\rho^{\gamma - 1} - b\rho^{\frac{1}{3}},
	\end{equation}
	
	where $a$ and $b$ here are function of $K, \gamma$ and the total mass of the Newtonian polytropic star. 
	
	Within our model, the total binding energy is initially given by Eq.\eqref{eqn: binding energy - eq 1} which, using the relation between $M$ and $R$ can be brought to the simple form of Eq.\eqref{eqn: binding energy - eq 2}. Indeed, if $\rho = \frac{3M}{4\pi R^3}$ is the total density of the star, we can rewrite Eq.\eqref{eqn: binding energy - eq 1} in the following way
	
	\begin{equation}\label{eqn: binding energy - eq 3}
		\begin{split}
			&\mathrm{E}_\text{bind} = a \rho^{\frac{1}{2}} - b\rho^{\frac{1}{3}}, \text{ where}\\
			& \hspace{-1.3cm} a = \frac{\hbar N_s^{\frac{4}{3}}}{8}\sqrt{\frac{4\pi G}{3}}, \hspace{2mm} b = \frac{3 N_s G M^{\frac{2}{3}}m}{2}  \sqrt[3]{\frac{4\pi}{3}}.
		\end{split}
	\end{equation}
	
	By comparing Eqs.(\ref{eqn: Newtonian E_bind}, \ref{eqn: binding energy - eq 3}), we see that our model fits in the Newtonian polytropic description with a $\gamma = \frac{3}{2}$. As $\frac{3}{2}>\gamma_\text{crit}$, the star is (classically) stable, with stability being achieved at the density value of:
	
	\begin{equation}
		\rho_* = \left(\frac{2b}{3a} \right)^6 = \frac{196608}{\pi}\left(\frac{m}{\hbar}\right)^6\frac{G^3 M^4}{N_s^2}.
	\end{equation}
	
	We can further express this explicitly as a function of $N_s$ and $m$ by making use of Eq.\eqref{eqn: M total mass - final form}.
	
	\begin{equation}
		\rho_* = \frac{196608}{\pi}\left(\frac{48}{71}\right)^4 \frac{c^6 M_p^4}{G^3 m^6}\frac{1}{N_s^{10/3}}\left(\sqrt{1+\frac{m^4}{M_p^4}N_s^{4/3}}-1\right)^4.
	\end{equation}
	
	This ensures that, even if arbitrarily close to the Schwarzschild radius, one can find a value for the total density which minimizes the binding energy, guaranteeing classical stability. 

    In principle, this analysis doesn't work for arbitrarily large $N_s$ at fixed $m$ since this case corresponds to the large $\d$ limit, in which $Z>\frac{1}{2}$ and the star lies inside the black hole horizon.


    \bibliographystyle{JHEP}      
    \bibliography{refs.bib}

@article{buchdahl1959general,
  author  = {Buchdahl, H. A.},
  title   = {General relativistic fluid spheres},
  journal = {Phys. Rev.},
  volume  = {116},
  pages   = {1027--1034},
  year    = {1959},
  doi     = {10.1103/PhysRev.116.1027}
}

@article{dadhich2025buchdahl,
  author  = {Dadhich, N. and Goswami, R. and Hansraj, C.},
  title   = {The Buchdahl bound denotes the geometrical Virial theorem},
  journal = {Ann. Phys.},
  volume  = {477},
  pages   = {169986},
  year    = {2025},
  doi     = {10.1016/j.aop.2025.169986},
  eprint  = {2304.10197},
  archivePrefix = {arXiv},
  primaryClass = {gr-qc}
}

@article{dadhich2020buchdahl,
  author  = {Dadhich, N.},
  title   = {Buchdahl compactness limit and gravitational field energy},
  journal = {J. Cosmol. Astropart. Phys.},
  volume  = {2020},
  number  = {04},
  pages   = {035},
  year    = {2020},
  doi     = {10.1088/1475-7516/2020/04/035},
  eprint  = {1903.03436},
  archivePrefix = {arXiv},
  primaryClass = {gr-qc}
}

@book{weinberg1972gravitation,
  author    = {Weinberg, S.},
  title     = {Gravitation and Cosmology: Principles and Applications of the General Theory of Relativity},
  publisher = {John Wiley \& Sons},
  address   = {New York},
  year      = {1972}
}

@article{boehmer2025buchdahl,
  author  = {Boehmer, C. G. and Dadhich, N. and Das, S.},
  title   = {Buchdahl stars and bounds with cosmological constant},
  year    = {2025},
  eprint  = {2507.00503},
  archivePrefix = {arXiv},
  primaryClass = {gr-qc}
}

@book{hartle2021gravity,
  author    = {Hartle, J. B.},
  title     = {Gravity: An Introduction to Einstein's General Relativity},
  publisher = {Cambridge University Press},
  address   = {Cambridge},
  year      = {2021}
}

@article{andreasson2008sharp,
  author  = {Andreasson, H.},
  title   = {Sharp bounds on $2m/r$ of general spherically symmetric static objects},
  journal = {J. Differ. Equ.},
  volume  = {245},
  pages   = {2243--2266},
  year    = {2008},
  doi     = {10.1016/j.jde.2008.05.010},
  eprint  = {gr-qc/0702137},
  archivePrefix = {arXiv},
  primaryClass = {gr-qc}
}

@article{michael2025,
  author  = {Saavedra, A. and Rubilar, G. and Fierro, O. and Gammon, M. and Mann, R. B.},
  title   = {Neutron stars in 4D Einstein-Gauss-Bonnet gravity},
  journal = {Phys. Rev. D},
  volume  = {111},
  pages   = {064071},
  year    = {2025},
  doi     = {10.1103/PhysRevD.111.064071},
  eprint  = {2412.15459},
  archivePrefix = {arXiv},
  primaryClass = {gr-qc}
}

@article{Zarro:2009gd,
  author  = {Zarro, C. A. D.},
  title   = {Buchdahl limit for d-dimensional spherical solutions with a cosmological constant},
  journal = {Gen. Rel. Grav.},
  volume  = {41},
  pages   = {453--468},
  year    = {2009},
  doi     = {10.1007/s10714-008-0675-8}
}

@article{Zarro:2009zz,
  author  = {Zarro, C. A. D.},
  title   = {Buchdahl and {T}olman-{O}ppenheimer-{V}olkoff limits for d-dimensional star-like solutions with a cosmological constant},
  journal = {Class. Quant. Grav.},
  volume  = {26},
  pages   = {035021},
  year    = {2009},
  doi     = {10.1088/0264-9381/26/3/035021}
}

@book{brack2018semiclassical,
  author    = {Brack, M. and Bhaduri, R.},
  title     = {Semiclassical Physics},
  publisher = {CRC Press},
  year      = {2018}
}

@book{Wald1984General,
  author    = {Wald, R. M.},
  title     = {General Relativity},
  publisher = {University of Chicago Press},
  address   = {Chicago},
  year      = {1984},
  doi       = {10.7208/chicago/9780226870373.001.0001}
}

@article{liebling2023dynamical,
  author  = {Liebling, S. L. and Palenzuela, C.},
  title   = {Dynamical boson stars},
  journal = {Living Rev. Relativ.},
  volume  = {26},
  pages   = {1},
  year    = {2023},
  doi     = {10.1007/s41114-023-00043-4}
}

@article{bovskovic2022soliton,
  author  = {Bo{\v{s}}kovi{\'c}, M. and Barausse, E.},
  title   = {Soliton boson stars, Q-balls and the causal Buchdahl bound},
  journal = {J. Cosmol. Astropart. Phys.},
  volume  = {2022},
  number  = {02},
  pages   = {032},
  year    = {2022},
  doi     = {10.1088/1475-7516/2022/02/032},
  eprint  = {2111.03870},
  archivePrefix = {arXiv},
  primaryClass = {gr-qc}
}

@article{alcubierre2022extreme,
  author  = {Alcubierre, M. and Barranco, J. and Bernal, A. and Degollado, J. C. and Diez-Tejedor, A. and Jaramillo, V. and Megevand, M. and N{\'u}{\~n}ez, D. and Sarbach, O.},
  title   = {Extreme $\ell$-boson stars},
  journal = {Class. Quant. Grav.},
  volume  = {39},
  pages   = {094001},
  year    = {2022},
  doi     = {10.1088/1361-6382/ac5fc2},
  eprint  = {2112.04529},
  archivePrefix = {arXiv},
  primaryClass = {gr-qc}
}

@article{alcubierre2023boson,
  author  = {Alcubierre, M. and Barranco, J. and Bernal, A. and Degollado, J. C. and Diez-Tejedor, A. and Megevand, M. and N{\'u}{\~n}ez, D. and Sarbach, O.},
  title   = {Boson stars and their relatives in semiclassical gravity},
  journal = {Phys. Rev. D},
  volume  = {107},
  pages   = {045017},
  year    = {2023},
  doi     = {10.1103/PhysRevD.107.045017},
  eprint  = {2212.02530},
  archivePrefix = {arXiv},
  primaryClass = {gr-qc}
}

@article{son2026vanishing,
  author  = {Son, E. J. and Kim, K. and Oh, J. J.},
  title   = {Vanishing compactness gap and fermionic compact dark matter in Ho{\v{r}}ava-Lifshitz gravity},
  journal = {Phys. Rev. D},
  volume  = {113},
  pages   = {104014},
  year    = {2026},
  doi     = {10.1103/zcwn-hq7g},
  eprint  = {2601.18079},
  archivePrefix = {arXiv},
  primaryClass = {gr-qc}
}

@article{mazur2004gravitational,
  author  = {Mazur, P. O. and Mottola, E.},
  title   = {Gravitational vacuum condensate stars},
  journal = {Proc. Nat. Acad. Sci.},
  volume  = {101},
  pages   = {9545--9550},
  year    = {2004},
  doi     = {10.1073/pnas.0402717101}
}

@article{jetzer1992boson,
  author  = {Jetzer, P.},
  title   = {Boson stars},
  journal = {Phys. Rep.},
  volume  = {220},
  pages   = {163--227},
  year    = {1992},
  doi     = {10.1016/0370-1573(92)90123-H}
}

@article{eby2016boson,
  author  = {Eby, J. and Kouvaris, C. and Nielsen, N. G. and Wijewardhana, L. C. R.},
  title   = {Boson stars from self-interacting dark matter},
  journal = {J. High Energy Phys.},
  volume  = {2016},
  number  = {02},
  pages   = {028},
  year    = {2016},
  doi     = {10.1007/JHEP02(2016)028},
  eprint  = {1511.04474},
  archivePrefix = {arXiv},
  primaryClass = {hep-ph}
}

@article{Tolman1939,
  author  = {Tolman, Richard C.},
  title   = {Static {S}olutions of {E}instein's {F}ield {E}quations for {S}pheres of {F}luid},
  journal = {Phys. Rev.},
  volume  = {55},
  pages   = {364--373},
  year    = {1939},
  doi     = {10.1103/PhysRev.55.364}
}

@article{OppenheimerVolkoff1939,
  author  = {Oppenheimer, J. R. and Volkoff, G. M.},
  title   = {On {M}assive {N}eutron {C}ores},
  journal = {Phys. Rev.},
  volume  = {55},
  pages   = {374--381},
  year    = {1939},
  doi     = {10.1103/PhysRev.55.374}
}
\end{document}